\documentclass{PoS}

\usepackage{amsmath}
\usepackage{amssymb}

\newcommand{\BR}{BR} 
\title{Charm CP: $\Delta A_{CP}$ and Radiative decays}

\ShortTitle{charmCP}

\author{\speaker{Amarjit Soni}\thanks{I want to thank Stefan Schacht for discussions. This research was supported in part by the U.S. DOE contract 
\#DE-SC0012704.}\\
        Physics Dept; Brookhaven National Lab, Upton, NY 11973; USA\\
        E-mail: \email{adlersoni@gmail.com}}

\abstract{Motivated by the very important discovery of CP violation in 
charm-decays for the first time, by the LHCb collaboration, the role of nearby resonances such as the scalar $f_0(1710)$ in accounting for the observed CP  is discussed. It is suggested that the influence of such a resonance may also explain the long-standing puzzle of such a large breaking of SU(3) seen in its decays. It is also explained that intervention of such resonance(s) will render first principles calculations of $\Delta A_{CP}$ rather difficult. Instead, it is proposed that searches for CP violation in simple radiative final states, such as $D^0 \to \gamma \phi [\to K^+ K^-]$, $D^0 \to l^+ l^- \pi$ etc.   has much better chance of theoretical precision studies and therefore theoretical and experimental investigations therein are strongly urged.}

\FullConference{37th International Symposium on Lattice Field Theory - Lattice2019\\
		16-22 June 2019\\
		Wuhan, China}

\begin{document}
\section{Introduction and Motivation}

Recently LHCb made an exciting discovery of direct CP violation in 
$D^0$  decays. 
Understanding CP violation is extremely important as naturalness arguments strongly suggest extensions of the SM should entail new CP-odd phases. Moreover, it seems very difficult to quantitatively understand the observed baryogenesis based on SM-CKM paradigm. To search for effects of new physics though requires comparison of SM prediction with experiment. Unfortunately,
this is usually a very difficult challenge for theory as direct CP violation 
entails non-perturbative effects. A famous example is the direct CP violation
parameter for $K \to \pi \pi$, $\epsilon'$ wherein after decades of effort significant progress in first principles calculation~\cite{Bai:2015nea, CK_LAT19}  is being made.

For addressing $D^0 \to h^+ h^-$ ($h = K, \pi$), these established methods
based on~\cite{LL01} are unlikely to be applicable primarily because $m_D$ is so much larger than $ 2 \times m_\pi$
and in fact when $D^0$ decays it can and therefore will readily go  to multi-particle final states.

One may be tempted to think that at $m_D$, SU(3) flavor symmetry [strictly speaking what is relevant is U-spin ($d-s$) symmetry] may be valid to a good approximation and result in a cancellation between multi-particle states but we suggest that this is unikely to be the case because of
intricate non-perturbative dynamics~\cite{CS84} as will be explained later and in fact, even more importantly, is exhibited in the experimental data~\cite{cPDG2018} on   $D^0$ decays.

In this work we make the following main points:

1) As explained in~\cite{AS0519} scalar and pseudoscalar resonances near $m_D$ play an important role in $\Delta A_{CP}$~\cite{LHCb_CP2019}  as well as in the decays of D-mesons.

2) Crude phenomenological estimates~\cite{AS0519} appear to suggest that the observed size of $\Delta A_{CP}$ with the influence of a neighboring resonance is roughly consistent with expectations from the SM but more accurate and reliable calculations are highly desirable. However, 
the presence of these resonances is likely to complicate even further first principles calculations of $\Delta_{ACP}$~\cite{LHCb_CP2019}

3) It is also suggested that these neighboring resonances are likely to be playing a role in significantly enhancing  the SU(3) breaking seen in $D^0$ decays. 

4) Attention is drawn to radiative decays of $D^0$, for example, $D^0 \to \gamma \rho^0 (\to \pi^+ \pi^-)$, $D^0 \to \gamma \phi (\to K^+ K^-)$~\cite{AS0519}\footnote{Work in progress with Stefan Schacht}. In contrast to $D^0$ decays to hadronic final states, these radiative channels are more amenable to lattice calculations. The photons are quite energetic and the remaining energy in the hadron system is around a GeV or less so current techniques for tackling LD issues ~\cite{DeltaMk2012} have a good chance of being applicable.

\section{LHCb observation of $\Delta A_{CP}$}

As is well known, for the past many years the LHCb collaboration has been trying to improve their search of direct CP asymmetries in $D^0 \to K^+ K^-$
and in $\pi^+ \pi^-$. In March 2019 at EW Moriond, they announced their important discovery~\cite{LHCb_CP2019},

\begin{eqnarray}
 \Delta A_{CP} \equiv [A_{CP}(K^+K^-)-A_{CP}(\pi^+\pi^-)] \nonumber \\
=(-15.4 \pm 2.9)X10^{-4}
\label{Eq:DeltaACP}
\end{eqnarray}

\noindent which represents  a 5.3 $\sigma$ signal of {\it direct} CP violation in charm decays. This result from LHCb is the first clear observation of CP violation (CPV) in the charm system and joins previous signals in
K and in B-systems.

\section{Back of the envelope estimate of $\Delta A_{CP}$}


A particularly  important process for CP violation is the $c \to u$ penguin graph,
especially wen the b-quark is in the penguin loop. Then, in the standard Wolfenstein parametrization~\cite{Wolf84} the $V_{ub}$ vertex is a very important source for the CP-odd phase in the CKM matrix. 

An extremely important point to understand is that in charm decays the penguin contribution is  extremely small
for several reasons. When the d and s are inside the penguin
then these tend to cancel to some degree as forced by the approximate orthogonality of the $2 \times 2$ sub-matrix due to the 1st two generations
and because of approximate SU(3) symmetry.  The b contribution is severely CKM-suppressed by the presence of
$V_{ub} \times V_{cb}^*$. For the singly Cabibbo suppressed modes such as 
$D^0 \to K^+ K^-$ that CKM-combination
should be compared with $\sin_{\theta_{Cabibbo}} \approx 0.225 \approx \lambda$, the Wolfenstein-parameter which enters the tree decay.  Since $V_{ub}$ goes as $\lambda^3$ and $V_{cb}$ as $\lambda^2$, the charm penguin is extremely small compared to the tree and $[penguin [P]/tree [T]] \approx \lambda^4 \approx 10^{-3}$ which is approximately the size of the expected CP-asymmetry in charm decays. However, by judiciously choosing some modes, for example, color suppressed modes~\cite{AS_PTEP13,NS_17}, one can get somewhat larger asymmetries by factors of order a few, $N = 3$.

Also, since 
partial rate asymmetry [PRA] defined as,

\begin{eqnarray}
\alpha_{PRA} =  (\BR [I \to F ] - \BR [\bar I \to \bar F])/ (\BR [I \to F ] + \BR [\bar I \to \bar F])
\label{Eq.PRA}
\end{eqnarray}

\noindent the PRA tends to go  $\alpha_{PRA} \propto [1/\sqrt(Br)]$ because $P$ is so small compared to $T$.

\section{Possible role of resonances in charm decays}

\subsection{Large SU(3) breaking in $D^0$ decays}

Resonances in the charm region can enter weak decays through the weak Hamiltonian and influence charm decays. Here,  in particular we want to draw attention to the scalar, $f_0$ with mass,
$m_{f_0} = 1723 MeV$ and width, $\Gamma_{f_0} = 139 MeV$\footnote{See PDGLIVE}. Note also that it has appreciable Br into $K^+ K^-$ and $\pi^+ \pi^-$,
$\approx 40\%$ and $\approx 16\%$.  

Now looking at Morningstar and Peardon's review~\cite{MP99} on glueball spectrum, it seems that $f_0$ at 1723 MeV is very likely rather rich in gluonia content.  If this interpretation is correct, then $f_0$ has a very interesting consequence. It was noted long ago~\cite{CS84} that the low-lying scalar and pseudoscalar glueballs couple to light quarks in proportion to their (constituent) mass. The gluonic interpretation of the $f_0$ would then also help understand the large breaking of SU(3) in its decays,

\begin{eqnarray}
Br [ f_0 \to K+ K^-]/[ f_0 \to \pi^+ \pi^- ]
\approx 2.5
\label{Eq.f0_SU3}
\end{eqnarray}.

In turn this breaking anticipated in~\cite{CS84} would help understand the large factor of  $\approx 2.8$ seen in $D^0$ decays to $K^+ K^-$ versus $\pi^+ \pi^-$\footnote{See also~\cite{hyc2012B}}.  It has been rather difficult to understand such a large SU(3) breaking in $D^0$ decays for a very long time. What factorization of tree amplitudes for these decays readily gives is that the breaking should go 
as the ratio of decay constants, $[f_k/f_{\pi}]^2 \approx 1.7$ which goes in the right direction but falls considerably short of the observed 2.8. But now with the factor of 2.5 originating above in gluon dynamics strongly suggests that this long standing mystery may now be resolved through $D^0-f_0$ mixing.

\subsection{ $f_0$ and $\Delta A_{CP}$}

It was suggested long ago~\cite{EHS91,AS94} that resonances can be used to estimate CP asymmetries. The main point is that in a channel which is dominated by a resonance,  the measured width of the resonance contains  the information about the rescattering phase that is needed for computing the CP-violating rate asymmetry in that channel.

\begin{figure}[h]
\centering
\includegraphics[width=0.5\textheight]{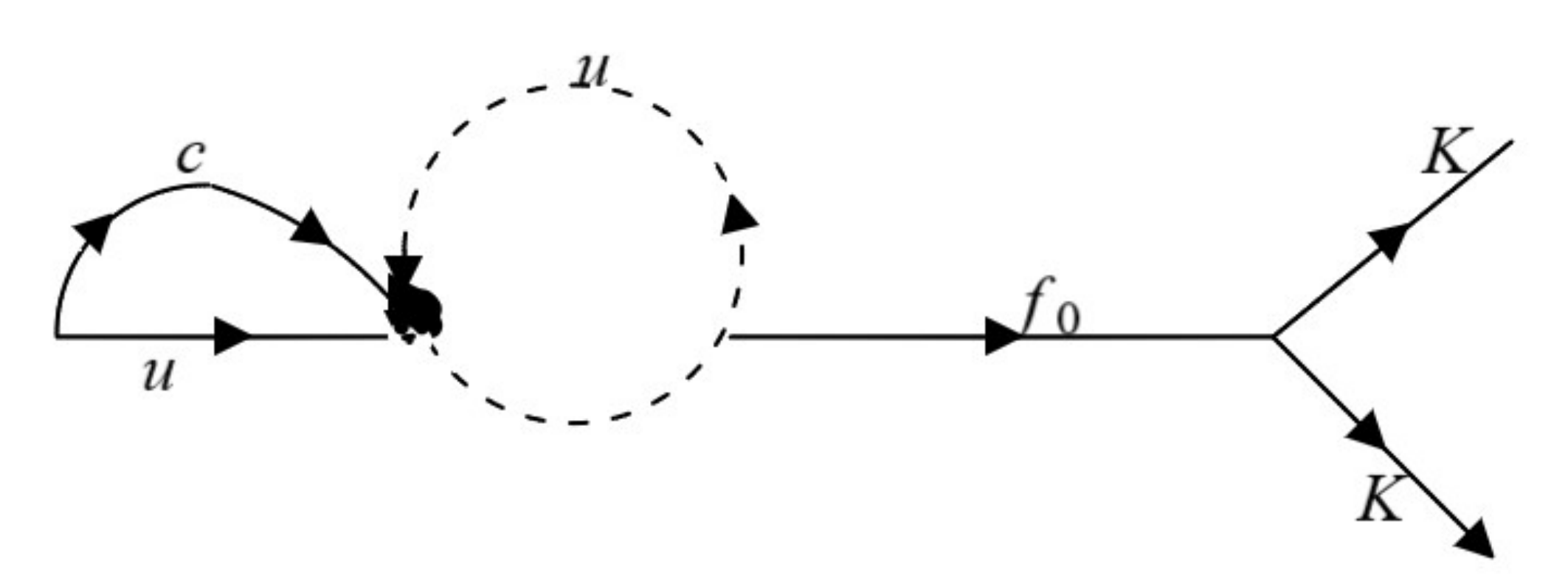}
\caption{$D^0$ decay proceeding to a final state such as $K^+$, $K^-$ (or $\pi^+ \pi^-$) via a  resonance {\it e.g} $f_0$; the weak 4-quark operator leads to the annihilation of the $D^0$ and the creation  of a u-quark pair which forms $f_0$.}
\label{fig:Feyn reson dash}
\end{figure}

We can use this idea for $D^0$ decays to $KK$ and $\pi\pi$ via the $f_0$. The key point to notice is that the 4-quark penguin operators have, in the conventional notation, $Q_5$ and $Q_6$, LXR operators which, on Fierz rearrangement contain $(S+P)\times(S-P)$ where $S=scalar$ and $P=pseudoscalar$.
Thereby, for instance, PS can annihilate a $D^0$ and create a $f_0$. The $f_0$ then propagates and travels for a brief instant and then decays to the final state. This is depicted in fig. 1. This leads to a rough estimate~\cite{AS0519}\footnote{See also~\cite{hnl2012}},

\begin{eqnarray}
\alpha_{KK} \approx 5.5 \times 10^{-4}
\label{alpha_KK}
\end{eqnarray}.
It is important to understand that while for simplicity, only the coupling of $f_0$ to $Q_6$ is emphasized, of course in a complete calculation coupling of all the operators in $H_{eff}$ and other neighboring scalar and pseudoscalar resonances should all be be taken into account.

Note also that there is a possibility that more nearby resonance(s) 
will be firmly 
established~\cite{BES2005,LHCb2014}\footnote{I want to thank Sheldon Stone for pointing out this higher resonance}
and then their effect may also need be included.

\subsection{CPT restrictions}

CPT theorem puts important constraint on PRAs since it requires that the total lifetime of particle and antiparticle must equal each other~\cite{HG87,ABES00,AS_PTEP13}
Thus PRAs in certain channels have to have opposite signs for the required cancellation to take place. As a simple illustration consider the quark level processes: $c \to s u \bar u$ versus $c \to d u \bar u$. CPT theorem requires that the PRAs in these two channels must cancel, so they have to have an opposite sign. 

As a result of this quark-level constraint we expect that the PRA for
$D^0 \to \pi^+  \pi^-$ must have an opposite sign to that for
$D^0 \to K^+  K^-$. So, as a consequence the two asymmetries entering into $\Delta A_{CP}$ add.

Another important consequence of CPT is that self-rescatterings (into the same final state) cannot generate PRAs. In other words, PRA for $c \to s u \bar u$ arises from on-shell scattering of $c \to d u \bar u$ and vice-versa.
But that restriction is only relevant for PRA, thus in final states with 3 or more particles in there  conjugate channels CP violating energy asymmeteries can arise and these can come about from self-rescattering effects.

Table~\ref{tab:CPT} summarizes our expectations of PRAs in the four prominent 2-body channels.

\begin{table}
\centering
\begin{tabular}{|c|c|c|c|}\hline
Mode &  BR ($10^{-3}$)  & Current PRA bound ($10^{-2}$) & Predicted in resonance model ($10^{-4}$) \\      
\hline
$K^{+} K^{-}$     & $3.97\pm.07$      & $- 0.07 \pm 0.11$ & $\approx$ 5.5 \\
$K_s K_s$         & $0.17 \pm 0.012$  & $-0.4 \pm 1.5$    & $\approx$ 13.2 \\
$\pi^{+} \pi^{-}$ & $1.407 \pm 0.025$ & $0.13 \pm 0.14$   & $\approx$ 5.8 \\
$\pi^{0} \pi^{0}$ & $0.822 \pm 0.025$ & $0.0 \pm 0.6$     & $\approx$ 7.5 \\
\hline
\end{tabular}
\caption{Br and CP asymmetries in the resonance model}
\label{tab:CPT}
\end{table}

\subsection {Resonance(s) influencing $\Delta A_{CP}$ and lattice calculations}

The Lellouch-Lusher~\cite{LL01} method for calculating finite volume correlation functions that is currently being used for $K \to \pi \pi$~\cite{Bai:2015nea} cannot directly be used for $D^0 \to KK (\pi\pi)$. Since $m_D$ is about 1.86 GeV, $D^0$ can readily decay to multiparticle final states which are rather challenging to handle on the lattice. To the extent that LHCb experiment has positive signal
for $\Delta A_{CP}$ which measures the {\it difference} in PRAs between $K^+ K^-$ and $\pi^+ \pi^-$ final states, one may think that flavor symmetry [Uspin] may provide a basis for
(approximate) cancellation between multiparticle states. But this is unlikely to hold since experimental data on $D^0$ decays shows large breaking of SU(3). Moreover, if the resonance hypothesis holds then there are even dynamical reasons to suggest such large breakings of SU(3).

\section {Candidates for precision testing the SM} 

Given that $D^0$ decays are in the region which is highly susceptible to long-distance non-perturbative effects it prompts one to think of simpler final states where first principles methods may have a better chance of success and also where experimental signals would not be too challenging. This thinking leads us
to suggest radiative decays of $D^0$, i.e. $D^0 \to \gamma + h_f$ where $h_f$
stands for simple final states such as $K^+K^-$, $\pi^+\pi^-$ etc. It may well be that the relevant hadronic final state is being influenced by nearby resonance(s). For example $\phi \to K^+ K^-$ and or $\rho \to \pi^+ \pi^-$. These radiative decays have attracted considerable phenomenological interest for a long time~\cite{KSW1997, BHS1993, BFS2001, IK2012, BH2017, BH2018, AHT2018}.
Study of radiative decays of $D^0$ may also reveal other resonances, for example, $\omega (1650)$, $\phi(1680)$, $\rho(1702)$~\cite{cPDG2018} etc. In passing, we note also that recently lattice methods are being able to address calculation of light meson scattering phases~\cite{PB2014,JLAB2016,JLAB2017,SM2018,CK_LAT19,DanH2019} in the $\approx$ GeV region.

\section {Summary}

It is suggested that scalar and pseudoscalar resonances near $m_D$ are playing an important role in the recently observed direct CP in charm decays by LHCb.
The presence of these resonances is likely to make first principle calculations
even harder. It is also suggested that these resonances are very likely playing an important role in the large observed SU(3) breaking. For first principle calculations, with an intent for precision tests of the Stndard Model, it may be better to focus on radiative decays such as
$D^0 \to \gamma K^+ K^-$, $\gamma \pi^+ \pi^-$ or $D^0 \to K (\pi) l^+ l^-$.

\bibliographystyle{JHEP}

\bibliography{refs4lat19}

\end{document}